\documentclass{article}
 \usepackage{inchmargs}

 \usepackage[hyphenbreaks]{breakurl}
 
\usepackage{hyperref}
\usepackage{cite}
\usepackage{amsmath,amssymb,amsfonts}
\usepackage{algorithmic}
\usepackage{graphicx}
\usepackage{textcomp}
\usepackage{xcolor}
\def\BibTeX{{\rm B\kern-.05em{\sc i\kern-.025em b}\kern-.08em T\kern-.1667em\lower.7ex\hbox{E}\kern-.125emX}}

\usepackage{booktabs} 
\usepackage{multirow} 
\usepackage{colortbl} 

\makeatletter
\renewcommand\footnoterule{%
 \kern-3\p@
 \hrule\@width 1\columnwidth
 \kern2.6\p@}
 \makeatother

\begin{document}


\title{Effective Wordle Heuristics}

\newcommand{\IEEEauthorblockA}[1]{ #1 }
\newcommand{\IEEEauthorblockN}[1]{ #1 }
\noindent
\author{
\IEEEauthorblockN{Ronald I. Greenberg}\\\IEEEauthorblockA{Loyola University Chicago\\Chicago, IL, USA\\rig@cs.luc.edu
}
}

\date{\today}
\maketitle

\pagestyle{myheadings}
\markboth{}{}

\begin{abstract}
While previous researchers have performed an exhaustive search to determine an optimal Wordle strategy, that computation is very time consuming and produced a strategy using words that are unfamiliar to most people.  With Wordle solutions being gradually eliminated (with a new puzzle each day and no reuse), an improved strategy could be generated each day, but the computation time makes a daily exhaustive search impractical.  This paper shows that simple heuristics allow for fast generation of effective strategies and that little is lost by guessing only words that are possible solution words rather than more obscure words.
\end{abstract}

\bigskip




\section{Introduction}

Wordle is a web-based word game that became a viral sensation in late 2021~\cite{Lum2021}.  Each day, a five-letter word is chosen that players attempt to guess within six tries.  After every guess, each letter is marked as either green, yellow or gray.  Green is used to mark a letter that is correct and in the correct position, and then yellow is used for each additional letter that appears in the answer but is not positioned correctly (once per occurrence in the answer), and the remaining letters of the guess are colored gray.  Discussions of the possible color responses conventionally use the denotations "G" for green, "Y" for yellow, and "B" for black (gray).

The game also has a "hard mode" option, which requires players to include letters marked as green and yellow in subsequent guesses. (The exact definition of hard mode does not mean that subsequent guesses have to be possible answers based
on the information obtained so far~\cite{Selby2022P1}, which might be a more intuitive definition that I suggest referring to as superhard mode.)

\subsection{Evolution of Word Lists and Notable Prior Results}

The original set of possible solutions was a list of 2315 words, while a larger list of 12,972 words was allowed for guesses~\cite{Selby2022P1}.  Much of prior analyses were done with these sets of words, and an exhaustive search was performed to obtain an optimal strategy (minimizing the average number of guesses across all possible solutions) beginning with the word "salet"~\cite{Selby2022P1}.  A similar optimal strategy was computed with a little bit smaller guess list of 10,657 words~\cite{BertsimasP2022}.  These results are included in Table~\ref{tab:prior-results}.

The New York Times Company purchased Wordle and began running it in February 2022 with a slightly altered solution list of 2309 words and later expanded the guess list with an additional 1881 words; these changes made little difference for prior analyses~\cite{Selby2022P2}. By July 2023, the list of possible solutions was transitioned to a list of 3158 words, making it much harder to perform exhaustive search for an optimal strategy, but such was done using parallel computation~\cite{Selby2023} as also shown in Table~\ref{tab:prior-results}.

The strategies mentioned so far use guess words that are unfamiliar to most people (e.g., "salet" and "tarse"), and this paper will show that resorting to such obscure guesses provides little gain in performance. The computations performed also were very time-consuming --- on the order of days (even with parallel computing in some cases).

Faster heuristic approaches have also been used; notable results also appear in Table~\ref{tab:prior-results}~\cite{Das2022,Luebeck2022}; they still use obscure guess words such as the start word "soare".

 \begin{table}
 \caption{Notable prior results}
 \label{tab:prior-results}
  \centering\vspace*{1ex}
  \begin{tabular}{lcclcc} \toprule
  {\bf method (\& start word)} &
  \shortstack{\bf available\\\bf guesses} &  \shortstack{\bf possible\\\bf solutions} & \shortstack{\bf average\\\bf guesses} &  \shortstack{\bf maximum\\\bf guesses} & \shortstack{\bf\% solved\\ \bf in $\leq6$\\\bf guesses} \\ \midrule
  {\bf optimal (salet)}~\cite{Selby2022P1} & 12,972 & 2,315 & 3.4212 & 5 & 100 \\ 
  {\bf optimal (salet)}~\cite{BertsimasP2022} & 10,657 & 2,315 & 3.42 & 5 & 100 \\
  {\bf optimal (tarse)~\cite{Selby2023}} & 14,885 & 3,158 & 3.5526 & 6 & 100 \\
  {\bf ad hoc heuristic (soare)~\cite{Das2022}} & 12,972 & 2,315 & 3.65 & 6 & 100 \\
  {\bf maximum entropy ... (soare)~\cite{Luebeck2022}} & 12,972 & 2,315 & 3.476 & 6 & 100 \\
  \bottomrule
  \end{tabular}
 \end{table}
 
\subsection{A New Direction}

This paper shows, however, that simple heuristics can obtain strategies with very good performance and without requiring use of obscure guess words; that is, the new results in this paper limit the guess words to possible solution words.  Some of the ideas for heuristics are not new, but this paper compares various heuristics in addition to limiting the guesses to more natural words.  The running time to generate a strategy with an appropriate heuristic is mere seconds, so the program can easily be rerun daily to take into account what words have already been used.  Many humans may be relying on some degree of recollection of past words, and this may provide a substantial performance gain as time goes on, since the game has not been reusing words.

\section{Results}

\subsection{Regular Mode}

Several heuristics were tested (along with some combinations); all are based on determining the set of possible secret words remaining after a given guess and organizing them in bins according to the color response that will be obtained if that secret word is the actual solution.  For example, if the first guess is "raise", there are 123 different bins; the largest with size 168 contains all those words from the solution list for which "raise" earns a response of "BBBBB".  Intuitively, the main thing we care about is just how many bins of each size result from a given guess.  But, especially when considering hard mode, it could also be desirable to measure and reduce the similarity among words in the same bin.  Experiments along that line were not particularly successful as pure strategies, but they provided some benefit as a tie-breaker in combination with a more conventional heuristic.  These ``conventional'' heuristics can be nicely cast into a framework of applying an $L^p$ norm (or ``metric'' more informally) to the bin sizes (generally summing the $p$-th power of each bin size), mostly agreeing with the description in~\cite{Rusin}. In any case, the goal was to find a simple heuristic (possibly with a secondary tie-break heuristic) that could be implemented efficiently and that would produce a low average number of guesses to find the solution (assuming all possible solutions are equally likely).  (In contrast, WordleBot~\cite{KatzC2023I2} assumes word likelihoods match the frequency of appearance of words in the New York Times.\footnote{In actual practice, the New York Times has been using words from the expanded solution list less frequently than words from the original solution list, so when employing heuristics on the expanded list, we use a list constructed with old words first so that, all else being equal, old words are chosen before new words. More generally, results do vary slightly with reordering of the word lists, but these effects are relatively minor.})

The heuristics tested included the following, listed here in order of best average number of guesses on the 2315-word solution list, and with correspondences to the framework of $L^p$ norms noted where applicable. All are expressed in terms of minimizing some measure calculated from the distribution of possible solutions into bins after trying a guess. In each case, the relevant heuristic was used to choose the starting word and then at each successive stage of the game. It should also be noted that in addition to only making guesses from the possible solution list, using a word consistent with (super)hard mode was preferred when there was a tie in the heuristic evaluation, and the search at a given stage was stopped any time that a bin distribution was found with all bins of size 1. There are also minor random effects based on the order of words in the list of solution candidates.
\begin{description}
 \item[negnumbins:] Minimize the negative of the number of bins. This corresponds to maximizing the $L^0$ norm.
 \item[negentropy:] Minimize the negative of the entropy. This corresponds to maximizing the the expression obtained from differentiating the $L^p$ norm at $p=1$, and is the basis of the WordleBot approach~\cite{KatzC2023I,KatzC2023I2}.
 \item[expbinsize:] Minimize the mathematical expectation of the size of the bin containing the solution. This corresponds to minimizing the $L^2$ norm.
 \item[Linfinty] This is an implementation of Rusin's description of the $L^{\infty}$ norm~\cite{Rusin}. According to the usual mathematical definition of the $L^{\infty}$ norm, we would just be minimizing the maximum bin size, but it is more effective to also include tie-break rules by considering how many bins have the maximum size, and then looking at how many have the next largest size, etc., as proposed by Rusin. This adds some complexity to the programming and is not as much of a pure, single heuristic as the others listed here, but we treat it here on a par with the other simple heuristics.
 \item[negnumsingletons:] Minimize the negative of the number of bins of size 1. This corresponds to the $L^{-\infty}$ norm.
 \item[maxsimilarity]: Minimize the maximum "similarity" among the words in a single bin. Similarity here is computed by distributing all the letters from the words in the bin into 26*5 new bins, each of which corresponds to a particular letter of the alphabet in a particular position of the guess word; then the negentropy for these 130 bins is computed. (Since different bins contain a different number of words, this strategy incorporates some element of the next.)
 \item[maxbinsize:] Minimize the maximum bin size. This is the classic $L^{\infty}$ norm that is a simplification of Rusin's $L^{\infty}$.
 \item[maxonediffs] Minimize the maximum number of word-pairs in a single bin for which the difference between the two words is a single letter in a single position.
 \end{description}

The results obtained from these heuristics are shown for the 2315-word solution list (ordered alphabetically) in the first section of Table~\ref{tab:heuristic-comparison} and for the 3158-word solution list (alphabetical except with the old words before the new words) in the second section. Each section orders the heuristics by average number of guesses. (There was little variation in the maximum number of guesses, and the heuristics with the best average number of guesses that could take 7 guesses only did so on a single word added to the solution list when it was expanded to 3158 words.)

 \begin{table}
 \caption{Comparison of heuristics as named above}
 \label{tab:heuristic-comparison}
  \centering\vspace*{1ex}
  \begin{tabular}{lcclcc} \toprule
  {\bf method (\& start word)} &
  \shortstack{\bf available\\\bf guesses} &  \shortstack{\bf possible\\\bf solutions} & \shortstack{\bf average\\\bf guesses} &  \shortstack{\bf maximum\\\bf guesses} & \shortstack{\bf\% solved\\ \bf in $\leq6$\\\bf guesses} \\ \midrule
  {\bf negnumbins (trace)} & 2,315 & 2,315 & 3.4600 & 6 & 100 \\
  {\bf negentropy (raise)} & 2,315 & 2,315 & 3.4955 & 6 & 100 \\
  {\bf expbinsize (raise)} & 2,315 & 2,315 & 3.5210 & 5 & 100 \\
  {\bf Linfinity (raise)} & 2,315 & 2,315 & 3.5564 & 5 & 100 \\
  {\bf negnumsingletons (brute)} & 2,315 & 2,315 & 3.5788 & 6 & 100 \\
  {\bf maxbinsize (arise)} & 2,315 & 2,315 & 3.5844 & 5 & 100 \\
  {\bf maxsimilarity (arise)} & 2,315 & 2,315 & 3.5901 & 5 & 100 \\
  {\bf maxonediffs (solar)} & 2,315 & 2,315 & 3.6695 & 6 & 100 \\
  \midrule
  {\bf negnumbins (caret)} & 3,158 & 3,158 & 3.6089 & 7 & 99.97 \\
  {\bf negentropy (raise)} & 3,158 & 3,158 & 3.6431 & 7 & 99.97 \\
  {\bf expbinsize (raise)} & 3,158 & 3,158 & 3.6602 & 6 & 100 \\
  {\bf Linfinity (arose)} & 3,158 & 3,158 & 3.7172 & 6 & 100 \\
  {\bf maxsimilarity (arose)} & 3,158 & 3,158 & 3.7394 & 6 & 100 \\
  {\bf maxbinsize (arose)} & 3,158 & 3,158 & 3.7565 & 6 & 100 \\
  {\bf negnumsingletons (bugle)} & 3,158 & 3,158 & 3.7977 & 7 & 99.81 \\
  {\bf maxonediffs (arose)} & 3,158 & 3,158 & 3.8046 & 7 & 99.87 \\
  \bottomrule
  \end{tabular}
 \end{table}

In addition, various combinations of heuristics were tried with one as primary and another for breaking ties. Individual heuristics that had the least average number of guesses were generally tried first, followed by another heuristic. Some best combinations are shown in Table~\ref{tab:combined-heuristics}.

 \begin{table}
 \caption{Good combination heuristics}
 \label{tab:combined-heuristics}
  \centering\vspace*{1ex}
  \begin{tabular}{lcclcc} \toprule
  {\bf method (\& start word)} &
  \shortstack{\bf available\\\bf guesses} &  \shortstack{\bf possible\\\bf solutions} & \shortstack{\bf average\\\bf guesses} &  \shortstack{\bf maximum\\\bf guesses} & \shortstack{\bf\% solved\\ \bf in $\leq6$\\\bf guesses} \\ \midrule
  {\bf negnumbins-expectation (trace)} & 2,315 & 2,315 & 3.4553 & 6 & 100 \\
  {\bf negnumbins-negentropy (trace)} & 2,315 & 2,315 & 3.4553 & 6 & 100 \\
  \midrule
  {\bf negnumbins-maxonediffs (caret)} & 3,158 & 3,158 & 3.6058 & 7 & 99.97 \\
  {\bf negnumbins-expectation (caret)} & 3,158 & 3,158 & 3.6067 & 7 & 99.97 \\
  {\bf negnumbins-maxbinsize (caret)} & 3,158 & 3,158 & 3.6067 & 7 & 99.97 \\
  {\bf negnumbins-negentropy (caret)} & 3,158 & 3,158 & 3.6067 & 7 & 99.97 \\
  {\bf negnumbins-maxsimilarity (caret)} & 3,158 & 3,158 & 3.6070 & 7 & 99.97 \\
  \bottomrule
  \end{tabular}
 \end{table}

 \subsection{Hard Mode}

Heuristics as discussed so far can have much weaker performance when required to play in hard (or superhard) mode. The greedy evaluation at each stage can overlook a lot that can be learned from exhaustive exploration of the game tree. Less attention was given to hard mode, but it had been hoped that heuristics like maxsimilarity and maxonediffs might be helpful for hard mode. In reality, the individual heuristics of Table~\ref{tab:heuristic-comparison} ranked mostly in the same order under superhard mode as for regular mode. As part of a combined heuristic, however, these approaches were helpful, and best strategies found for superhard mode are as indicated in Table~\ref{tab:hard-mode}

 \begin{table}
 \caption{Good combination heuristics for (super)hard mode}
 \label{tab:hard-mode}
  \centering\vspace*{1ex}
  \begin{tabular}{lcclcc} \toprule
  {\bf method (\& start word)} &
  \shortstack{\bf available\\\bf guesses} &  \shortstack{\bf possible\\\bf solutions} & \shortstack{\bf average\\\bf guesses} &  \shortstack{\bf maximum\\\bf guesses} & \shortstack{\bf\% solved\\ \bf in $\leq6$\\\bf guesses} \\ \midrule
  {\bf negnumbins-maxonediffs (trace)} & 2,315 & 2,315 & 3.5322 & 8 & 99.65 \\
  {\bf negnumbins-similarity (trace)} & 2,315 & 2,315 & 3.5322 & 8 & 99.65 \\
  {\bf negnumbins-expectation (trace)} & 2,315 & 2,315 & 3.5335 & 8 & 99.65 \\
  {\bf negnumbins-maxbinsize (trace)} & 2,315 & 2,315 & 3.5335 & 8 & 99.65 \\
  {\bf negnumbins-negentropy (trace)} & 2,315 & 2,315 & 3.5335 & 8 & 99.65 \\
  \midrule
  {\bf negnumbins-maxonediffs (caret} & 3,158 & 3,158 & 3.7283 & 9 & 98.73 \\
  {\bf negnumbins-maxsimilarity (caret)} & 3,158 & 3,158 & 3.7305 & 9 & 98.64 \\
  {\bf negnumbins-negentropy (caret)} & 3,158 & 3,158 & 3.7324 & 9 & 98.67 \\
  \bottomrule
  \end{tabular}
 \end{table}

 These heuristics are good in terms of the average number of guesses but leave something to be desired in terms of the maximum number of guesses. Though each of the cases in Table~\ref{tab:hard-mode} with a maximum of 9 guesses only requires 9 guesses on a single word that was added when the solution list was expanded to 3158 words, all the results in Table~\ref{tab:hard-mode} are quite far from an optimal strategy; for the 2315-word solution list, there is an optimal superhard-mode strategy starting with "scamp" that requires a maximum of 5 guesses and averages 3.7162 guesses. (This strategy was found using the same code base as for testing the heuristics reported in this paper. The exhaustive search becomes more feasible when limiting to hard mode, but without using a particularly powerful computer or delving into techniques to prune the search tree, the program ran for about three weeks, and the longer 3158-word solution list would present greater challenge.)

\subsection{Accounting for Previously Used Words}

Settling on the negnumbins heuristic with the expectation tie-breaker as generally a fast-running and effective approach in regular mode, this approach was applied to each of the past daily scenarios, taking into account words that have already been used and assuming they would not be reused. Though the list of possible solutions has evolved over time, this experiment was done using the current 3158-word list and eliminating, day by day, the actually used past words. (There have been a few occasions when some users were mistakenly presented with an alternative word~\cite{Molina2022}, but the intended word was taken as the actual word for the day.)

As of August 12, 2024, using the chosen heuristic on all remaining possible solutions, the average number of guesses (assuming all words to be equally likely and allowing all 3158 words to be used as guesses) has gone down only a little to 3.4905; the maximum number of guesses remains at 6 but only for nine solution words, all of which were not in the original list of 2315. The result is almost as good if the used words are eliminated not only as possible solutions but as possible guesses; then the average number of guesses is 3.5030 with a maximum of 6 guesses on just seven solution words, all of which were not in the original list of 2315.

The words added to the solution list by the New York Times are being used very rarely, however (only six actually used to date). If these extra words are eliminated from consideration as possible solutions, the heuristic achieves an average of 3.2625 guesses and a maximum of 5.

The NYT editors may soon want to consider making more choices from solutions that were added in the expansion to 3158 possible solutions. Or they may start reusing words; in the mean time, using the heuristic recommended in this paper, a new recommended strategy is being made available daily at \url{http://rig.cs.luc.edu/~rig/wordle}. These strategies allow for the possibility of solutions being chosen from the full 3158-word list; still they use the previously noted approach of preferring guesses from the original 2315-word list when such a guess seems heuristically to be as good as other guesses. The code used is also available there.

\section{Conclusion}

We have compared heuristics for solving Wordle and have found a fast heuristic to generate a good Wordle strategy each day (using only guesses from the 3158-word possible solution list), taking into account past history (assuming solutions are not reused). Surprisingly, the theoretically sound strategy of making guesses to maximize entropy is not quite best in practice (assuming all allowed solutions are equally likely). Instead, it is generally best to maximize the number of bins into which solutions divide based on the possible differing color responses, with ties being broken by maximizing entropy or by minimizing expected bin size. (The latter of these tie breakers is a slightly faster approach, since it avoids floating-point arithmetic.)

For (super)hard mode, it has been shown that the best results may be obtained with some other combination heuristics using more information than just the number of words per bin. These hard-mode results have good average numbers of guesses but have difficulty matching the maximum number of guesses from strategies found through exhaustive search.

The results reported in this paper seem fairly resilient to the exact list of possible solutions, but future work might investigate whether there is something anomalous about working with 5-letter English words or the conclusions here are more generalizable. Future work also might seek a better heuristic approach for hard mode or perhaps a different way of combining heuristics; for example, it might make sense to use a certain heuristic for the initial guesses and a different heuristic for later guesses.

\section{Postscript}

As of February 2, 2026, the New York Times is starting to allow the possibility of reusing past words and actually did so on that day.  As of February 11, however, they have only reused a word once.  So the heuristics discussed in this paper still seem to be useful in practice at this time.

\bibliographystyle{unsrt} 
\bibliography{refs}

\end{document}